\documentclass{article}
\begin{document}

\title{Onsager algebra and cluster XY-models in a transverse magnetic field}

\author{Jacques H.H.\ Perk\\
Department of Physics, Oklahoma State University\\
Stillwater, OK 74078-3072, USA}
\maketitle

\begin{abstract}
The correlation functions of certain $n$-cluster XY models are explicitly
expressed in terms of those of the standard Ising chain in transverse field. 
\end{abstract}

\newcommand{\be}{\begin{equation}}
\newcommand{\ee}{\end{equation}}
\newcommand{\ba}{\begin{eqnarray}}
\newcommand{\ea}{\end{eqnarray}}
\newcommand{\rme}{\mathrm{e}}
\newcommand{\rmi}{\mathrm{i}}
\newcommand{\rmc}{\mathrm{c}}
\newcommand{\rbp}{\mathbf{P}}
\newcommand{\rbq}{\mathbf{Q}}
\newcommand{\mch}{\mathcal{H}}
\newcommand{\mco}{\mathcal{O}}
\newcommand{\tr}{\mathrm{Tr}\,}

\section{Introduction}

We start with the Hamiltonian, eq.~(1) in \cite{ZG},
\be
\mch=-J\sum_{j=1}^N\sigma^x_j\bigg(\prod_{k=j+1}^{j+n}\sigma^z_k\bigg)
\sigma^x_{j+n+1}-H\sum_{j=1}^N\sigma^z_j,\quad N\equiv(n+1)N_1,
\label{ham}
\ee
with periodic boundary conditions,
$\sigma^{\alpha}_{j+N}\equiv\sigma^{\alpha}_j$,
for $\alpha=x,y,z$. In this note we are interested in the factorization
of certain correlations functions in the bulk thermodynamic limit
$N\to\infty$. The calculation is easiest, if we use periodic boundary conditions and chain length $N\equiv0$ mod $n+1$.%
\footnote{If $N\not\equiv0$ mod $n+1$, we can modify the boundary
conditions far away from the operators of interest, such that the
factorization still works. For the case of finite chains with
open boundary conditions the factorization can be seen to be exact
for finite $N$ also.}

Hamiltonian (\ref{ham}) is a special case of the generalized XY-model
discussed by Suzuki in the early 1970s \cite{Suzuki1,Suzuki2}.
The zero-field XY model with isotropic interactions has already been
introduced by Nambu in 1950 \cite{Nambu}. A more detailed study with
anisotropic interaction was done by Lieb, Schultz and Mattis \cite{LSM},
while Katsura \cite{Katsura} studied the thermodynamic properties in
a magnetic field. The special case of (\ref{ham}) with $n=0$ is called the
Ising chain in transverse field and was treated in more detail by
Pfeuty \cite{Pfeuty}.

The case of (\ref{ham}) with $n=1$ has been studied by many authors and can
be shown to be equivalent to the zero-field XY model of \cite{Nambu,LSM},
using a duality transform. This is already to be expected from the
Onsager algebra (60) and (61) in \cite{Onsager}: The zero-field
Hamiltonian of \cite{Nambu,LSM} is a linear combination of $A_1$ and
$A_{-1}$, whereas (\ref{ham}) with $n=1$ is a linear combination of $A_2$
and $A_0$. Expressing the $A_k$ in terms of Kaufman's Gamma operators
\cite{Kaufman}, the map is then just a shift of all
$\Gamma_{2j-1}\to\Gamma_{2j+1}$, keeping the $\Gamma_{2j}$ fixed, or
equivalently $\rbp_j\to\rbp_{j+1}$,
$\rbq_j\to\rbq_{j}$ at the beginning of section 3 of \cite{Kaufman}.

Looking at Nambu's figure 1 \cite{Nambu} one may already get the idea
that his XY Hamiltonian (8) splits into two commuting Hamiltonians. Indeed,
in section 6 of \cite{PC} it is worked out in detail how the more
general alternating zero-field XY Hamiltonian splits into two
transverse-field Ising chain Hamiltonians that commute, implying
the factorization of correlation functions.%
\footnote{We have used similar factorizations in several other papers,
see for example section 7 of \cite{CP}, eq.~(3.11) of \cite{PCS},
below eq.~(7) in \cite{P}, eq.~(58) of \cite{MPS}, and eq.~(22) of
\cite{APdimer}. Similar to figure 1 in \cite{Nambu} for  case $n=0$,
the factorization for case $n=1$ is also implicitly present in figure 3
of \cite{SAFFFPV}, which used $\check a_k$ for what Kaufman called
$\Gamma_k$. These factorizations closely parallel related factorizations
in 2D classical spin models, see e.g.\ section 10.3 of \cite{BaxterBook}.
This is not surprizing, as such relationships between
$d$-dimensional quantum systems and $(d+1)$-dimensional classical
systems was particularly advertised by Suzuki \cite{Suzuki3}.}
This means, more generally for the case $n>0$, that we can also
calculate the correlation functions studied in \cite{ZG} in terms
of the known results for the transverse-field Ising chain given
in \cite{AP} and references cited there.

We will show this next in section 2. In Section 3 we shall discuss
the more general situation using Onsager's algebra \cite{Onsager}.
We close with a conclusion in section 4.

%%%%%%%%%%%%%%%%%%%%%%%%%%%%%%%%%%%%%%%%%%%%%%%

\section{Fermionization and factorization}

Following Kaufman's spinor analysis \cite{Kaufman}, we introduce
Clifford algebra operators through the Jordan--Wigner
transformation \cite{JW},
\be
\Gamma_{2j-1}=\bigg(\prod_{k=1}^{j-1}\sigma^z_k\bigg)\sigma^x_j
=\rbp_j,\quad
\Gamma_{2j}=\bigg(\prod_{k=1}^{j-1}\sigma^z_k\bigg)\sigma^y_j
=\rbq_j,\quad
\sigma^z_j=-\rmi\Gamma_{2j-1}\Gamma_{2j},
\label{Gamma}\ee
satisfying
\be
\Gamma_k\Gamma_l+\Gamma_l\Gamma_k=2\delta_{kl}\mathbf{1}.
\ee
This is, in fact, eq.~(15) with eq.~(6) in \cite{Kaufman}, omitting the
asterisks there. Eq.~(\ref{Gamma}) does not appear explicitly as such in \cite{JW}; Kaufman took it from eq.~(9) in \cite{BW} instead.

Equivalently, following Jordan and Wigner \cite{JW}, Nambu \cite{Nambu}
and Lieb, Schultz and Mattis \cite{LSM}, we could have used the fermion
creation and annihilation operators,
\be
c^{\vphantom{\dagger}}_j=\frac12(\Gamma_{2j-1}-\rmi\Gamma_{2j}),\quad
c^{\dagger}_j=\frac12(\Gamma_{2j-1}+\rmi\Gamma_{2j}),
\ee
but that is less convenient for our purpose. We note that in \cite{Nambu}
the $\Gamma_k$ have been written as $x_k$ and that in \cite{LSM} the
$\rbp_j$ and $\rbq_j$ have been called $A_j$ and
$-\rmi B_j$. In \cite{PC} we used $\gamma_j=\Gamma_j/\sqrt{2}$.
More recently these operators are also called Majorana fermions in
reference to \cite{Majorana}. However, these operators appeared
already as eq.~(I) on p.~650 in \cite{JW}, identifying
$\rbp_j=\alpha_j$ and $\rbq_j=\alpha_{N+j}$.

As a result, the Hamiltonian becomes
\be
\mch=\sum_{j=1}^N\Big[\rmi J\Gamma_{2j}\Gamma_{2j+2n+1}+
\rmi H\Gamma_{2j-1}\Gamma_{2j}\Big],
\ee
up to a boundary term that we can ignore in the thermodynamic limit
for the quantities we discuss in this paper. This is so, as long as
we stay with operators in the ``even sector'' with $\Gamma_j$ operators
clearly grouped in pairs. It fails if the ``odd sector'' becomes
important, see e.g.~\cite{PC,MBA}. This complication does not show up in
the open boundary case, but then one has to deal with boundary effects.

Similar to eq.~(6.8) in \cite{PC}, we can next relabel the operators
according to
\be
\Gamma^{(p)}_{2k+1}=\Gamma_{2p+2k(n+1)+1},\quad
\Gamma^{(p)}_{2k+2}=\Gamma_{2p+2k(n+1)+2},
\label{tra}
\ee
for $p=0,\cdots,n$, $k=0,\cdots,N_1-1$, and satisfying
\be
\Gamma^{(p)}_k\Gamma^{(q)}_l+\Gamma^{(q)}_l\Gamma^{(p)}_k=
2\delta_{pq}\delta_{kl}\mathbf{1}.
\ee
For given $\Gamma_j$, we can find the $p$ and $k$ in (\ref{tra}) using
\be
p=\left\lfloor(n+1)\left\{\frac{j-1}{2(n+1)}\right\}\right\rfloor,\quad
k=\left\lfloor\frac{j-1}{2(n+1)}\right\rfloor,
\label{map}
\ee
where $\lfloor x\rfloor$ stands for the floor or integer part of $x$ and
$\{x\}$ is the fractional part of $x$. The extra $+1$ or $+2$ in
(\ref{tra}) corresponds to $j$ being odd or even. We find
\be
\mch=\sum_{p=0}^n\mch^{(p)},\quad\mch^{(p)}=
\sum_{k=1}^{N_1}\Big[\rmi J\Gamma^{(p)}_{2k}\Gamma^{(p)}_{2k+1}+
\rmi H\Gamma^{(p)}_{2k-1}\Gamma^{(p)}_{2k}\Big].
\label{factored}
\ee
We can now define
\ba
&\sigma^{z(p)}_j=-\rmi\Gamma^{(p)}_{2j-1}\Gamma^{(p)}_{2j},
\hspace{26pt}&\\
&\displaystyle\sigma^{x(p)}_j=
\bigg(\prod_{k=1}^{j-1}\sigma^{z(p)}_k\bigg)\Gamma^{(p)}_{2j-1},&\quad
\sigma^{y(p)}_j=
\bigg(\prod_{k=1}^{j-1}\sigma^{z(p)}_k\bigg)\Gamma^{(p)}_{2j},
\ea
so that
\be
\mch^{(p)}=
-J\sum_{j=1}^{N_1}\sigma^{x(p)}_j\sigma^{x(p)}_{j+1}
-H\sum_{j=1}^{N_1}\sigma^{z(p)}_j,\quad p=0,\cdots,n.
\label{Ising}
\ee
Thus $\mch$ is decomposed into $n+1$ commuting Ising chains in transverse field, with identical coupling $J$ and field $H$ and factorizing
$\exp(\beta\mch)$, as we can again ignore the boundary effect in the
large $N_1$ limit. This causes the partition function and the spin
correlations to factorize in the thermodynamic limit.%
\footnote{For finite $N$ the spin correlations for the system
(\ref{ham}), with periodic boundary conditions and $N=N_1(n+1)$,
become ratios of sums with four factorized terms, cf.\
\cite[eqs. (35), (39)]{Kaufman}. If we had applied open boundary
conditions, the correlations would simply factorize, even for
$N\not\equiv0$ mod $n+1$.}

%%%%%%%%%%%%%%%%%%%%%%%%%%%%%%%%%%%%%%%%%%%%%%%

Let us now consider the equilibrium pair correlation functions
\be
X^{(\rmc)}(k)=\langle\sigma^x_{j}\sigma^x_{j+k}\rangle,\quad
Y^{(\rmc)}(k)=\langle\sigma^y_{j}\sigma^y_{j+k}\rangle,\quad
Z^{(\rmc)}(k)=\langle\sigma^z_{j}\sigma^z_{j+k}\rangle,
\ee
for cluster model (\ref{ham}) in the large $N$ limit.%
\footnote{Using the methods in e.g.\ \cite{PC,CP}, what follows can
also be generalized to time-dependent correlations
$\langle\sigma^{\lambda}_{j}(t)\sigma^{\mu}_{j+k}\rangle$
with $A(t)\equiv\rme^{\rmi\mch t}A\rme^{-\rmi\mch t}$.}
Here $\langle O\rangle$ stands for either the ground state expectation
of $O$, or the thermal expectation
$\langle O\rangle=\tr O\,\rme^{-\beta\mch}/\tr\rme^{-\beta\mch}$. Now
\ba
\sigma^x_{j}\sigma^x_{j+k}&=&-\rmi\Gamma_{2j}\bigg(\prod_{l=j+1}^{j+k-1}
(-\rmi\Gamma_{2l-1}\Gamma_{2l})\bigg)\Gamma_{2j+2k-1},
\label{xx}\\
\sigma^y_{j}\sigma^y_{j+k}&=&-\rmi\Gamma_{2j-1}\bigg(\prod_{l=j+1}^{j+k-1}
(-\rmi\Gamma_{2l-1}\Gamma_{2l})\bigg)\Gamma_{2j+2k},
\label{yy}\\
\sigma^z_{j}\sigma^z_{j+k}&=&(-\rmi\Gamma_{2j-1}\Gamma_{2j})
(-\rmi\Gamma_{2j+2k-1}\Gamma_{2j+2k}).
\label{zz}
\ea
Therefore, we immediately conclude that
\be
X^{(\rmc)}(k)=Y^{(\rmc)}(k)=0,\quad\mbox{if }k\not\equiv0\mbox{ mod }n+1,
\label{xy0}
\ee
as then $\Gamma_{2j}$ and $\Gamma_{2j+2k-1}$, (and similarly
$\Gamma_{2j-1}$ and $\Gamma_{2j+2k}$), belong to different $p$ values,
causing odd numbers of $\Gamma_l$ to fall into the two corresponding
$\mch^{(p)}$.

Let us introduce
\be
X(k)=\langle\sigma^x_{j}\sigma^x_{j+k}\rangle,\quad
Y(k)=\langle\sigma^y_{j}\sigma^y_{j+k}\rangle,\quad
Z(k)=\langle\sigma^z_{j}\sigma^z_{j+k}\rangle,
\ee
for pair correlations in the Ising chain in transverse field with
coupling $J$ and field $H$, and
\be
X^{\ast}(k)=\langle\sigma^x_{j}\sigma^x_{j+k}\rangle,\quad
Y^{\ast}(k)=\langle\sigma^y_{j}\sigma^y_{j+k}\rangle,\quad
Z^{\ast}(k)=\langle\sigma^z_{j}\sigma^z_{j+k}\rangle,
\ee
for the dual case with coupling $H$ and field $J$, and obtained after the
duality transform $\Gamma_l\to\Gamma_{l-1}$, so that
$\sigma_j^z=\sigma_j^{x\ast}\sigma_{j+1}^{x\ast}$ and
$\sigma^x_{j}\sigma^x_{j+1}=\sigma_{j+1}^{z\ast}$, see
\cite[p.~123]{Onsager} and \cite[p.~1237]{Kaufman}.

If now we set $j=1$ and replace $k$ by $k(n+1)$ in (\ref{xx}), we can
rewrite
\ba
\sigma^x_{1}\sigma^x_{1+k(n+1)}&=&-\rmi\Gamma^{(0)}_2\bigg(\prod_{l=2}^{k}
(-\rmi\Gamma^{(0)}_{2l-1}\Gamma^{(0)}_{2l})\bigg)\Gamma^{(0)}_{2k+1}
\nonumber\\
&&\times\prod_{p=1}^n\bigg[-\rmi\Gamma^{(p)}_{1}\bigg(\prod_{l=2}^{k}
(-\rmi\Gamma^{(p)}_{2l-2}\Gamma^{(p)}_{2l-1})\bigg)\Gamma^{(p)}_{2k}\bigg],
\ea
and an analogous expression for (\ref{yy}). Similar expressions are found
for other values of $j$, but we really only need the result for $j=1$ as
Hamiltonian (\ref{ham}) is translationally invariant.
Hence, we find the factorizations
\be
X^{(\rmc)}\big(k(n+1)\big) = X(k) X^{\ast}(k)^n,\quad
Y^{(\rmc)}\big(k(n+1)\big) = Y(k) X^{\ast}(k)^n.
\label{xy}
\ee
From (\ref{zz}) and using (\ref{map}) we see that
\be
\sigma^z_{j}\sigma^z_{j+k}=\sigma^{z(p_1)}_{k_1}\sigma^{z(p_2)}_{k_2},
\ee
with $p_1=p_2$ only if $k$ is a multiple of $n+1$. Therefore, we find
\be
Z^{(\rmc)}\big(k(n+1)\big) = Z(k),\quad\mbox{but }
Z^{(\rmc)}(m)=M_z^2, \mbox{ if }m\not\equiv0 \mbox{ mod }n+1,
\label{zz0}
\ee
where $M_z=\langle\sigma^z_j\rangle$ is the $z$-magnetization in the
Ising chain (\ref{Ising}). Now we have only one or two factors remaining,
as the other $n$ or $n-1$ factors are trivially equal to one.

%%%%%%%%%%%%%%%%%%%%%%%%%%%%%%%%%%%%%%%%%%%%%%%

Next we consider the ``well-tailored'' cluster operators in eq.~(24)
of \cite{ZG},
\ba
&&\mco^{(n)}_j=\bigg(\prod_{k=1}^{j-1}\sigma^z_k\bigg)
\bigg(\prod_{k=0}^{\lfloor n/2\rfloor}
\sigma^{y}_{j+2k}\sigma^{\vphantom{y}x}_{j+2k+1}\bigg),\quad\mbox{if $n$ is odd},
\nonumber\\
&&\mco^{(n)}_j=\sigma^{\vphantom{y}x}_j
\bigg(\prod_{k=1}^{\lfloor n/2\rfloor}\sigma^y_{j+2k-1}\sigma^{\vphantom{y}x}_{j+2k}\bigg),\quad\mbox{if $n$ is even}.
\ea
Here, compared to \cite{ZG}, we shifted the $j$ by $n$ for the odd case,
in order to get a more uniform result after Jordan--Wigner transform
(\ref{Gamma}). More precisely, using
$\sigma^y_k\sigma^{\vphantom{y}x}_{k+1}=\rmi\Gamma_{2k-1}\Gamma_{2k}$,
we obtain, both for $n$ odd and for $n$ even,
\be
\mco^{(n)}_j=\rmi^{\lfloor(n+1)/2\rfloor}
\bigg(\prod_{k=1}^{j-1}(-\rmi\Gamma_{2k-1}\Gamma_{2k})\bigg)
\prod_{l=0}^n\Gamma_{2j+2l-1}.
\ee
From this we find, using $(-1)^{\lfloor(n+1)/2\rfloor}=(-1)^{n(n+1)/2}$,
\ba
\mco^{(n)}_j\mco^{(n)}_{j+r}&=&(-1)^{n(n+1)/2}(-\rmi)^{n+1}
\prod_{l=0}^n\Gamma_{2j+2l}
\nonumber\\
&&\times
\bigg(\prod_{k=j+n+1}^{j+r-1}(-\rmi\Gamma_{2k-1}\Gamma_{2k})\bigg)
\prod_{l=0}^n\Gamma_{2j+2r+2l-1}.
\ea
Next we use the relabeling (\ref{tra}) and reorder the $\Gamma^{(p)}$'s
by increasing $p$. That costs exactly $\frac12n(n+1)$ minus signs. We can
set $j=1$, as the correlation to be gotten cannot depend on $j$.
Thus we get
\be
\mco^{(n)}_1\mco^{(n)}_{r+1}=\prod_{p=0}^n
\prod_{k=1}^{\lfloor(r+n-p)/(n+1)\rfloor}
(-\rmi\Gamma^{(p)}_{2k}\Gamma^{(p)}_{2k+1}).
\ee
Hence, we find
\be
\langle\mco^{(n)}_j\mco^{(n)}_{j+r}\rangle=
\prod_{p=0}^n X\Big(\Big\lfloor\frac{r+p}{n+1}\Big\rfloor\Big),
\ee
with $X(k)=\langle\sigma^x_{j}\sigma^x_{j+k}\rangle$ for the Ising chain (\ref{Ising}) and $\lfloor x\rfloor$ the integer part
of $x$.

As $r\to\infty$, the above result goes to $(\langle\sigma^x_{j}\rangle)^{2(n+1)}$,
the $2(n+1)$ power of the Ising order parameter $m_x=\langle\sigma^x_{j}\rangle$.

%%%%%%%%%%%%%%%%%%%%%%%%%%%%%%%%%%%%%%%%%%%%%%%

\section{Onsager algebra for Ising model}

In his solution of the 2-dimensional Ising model, Onsager
introduced \cite{Onsager}%
\footnote{The comparison with \cite{Onsager} requires the
identification $s_j=\sigma^x_j$ and $C_j=\sigma^z_j$, while using
a rotated representation of the Pauli matrices, i.e.
$\sigma^x_j\leftrightarrow\sigma^z_j$, $\sigma^y_j\to-\sigma^y_j$.}
\ba
&&A_n=\sum_{j=1}^N \sigma^x_j
\bigg(\prod_{k=j+1}^{j+n-1}\sigma^z_k\bigg)\sigma^x_{j+n},
\\
&&G_n=\frac12\mathrm{i}\sum_{j=1}^N \bigg[\sigma^x_j
\bigg(\prod_{k=j+1}^{j+n-1}\sigma^z_k\bigg)\sigma^y_{j+n}
+\sigma^y_j
\bigg(\prod_{k=j+1}^{j+n-1}\sigma^z_k\bigg)\sigma^x_{j+n}\bigg].
\ea
We have to assume periodicity $\sigma^\alpha_{j\pm N}=\sigma^\alpha_{j}$,
$\alpha=x,y,z$. In addition, as $(\sigma^z_k)^2=1$, we have
\be
\prod_{k=j+1}^j\sigma^z_k=1,\quad
\prod_{k=j+1}^{j-m}\sigma^z_k=\prod_{k=j-m+1}^{j}\sigma^z_k,
\ee
so that, using this and the Pauli matrix product rules, we find
\ba
&&A_0=-\sum_{j=1}^N \sigma^z_j,\quad A_{-n}=\sum_{j=1}^N \sigma^y_j
\bigg(\prod_{k=j+1}^{j+n-1}\sigma^z_k\bigg)\sigma^y_{j+n},\\
&&A_{n\pm N}=-P A_n=-A_nP,\qquad P\equiv
\prod_{k=1}^N\sigma^z_k,\label{P}\\
&&G_0=0,\quad G_{-n}=-G_n,\quad G_{n\pm N}=-P G_n=-G_nP,\\
&&A_{n\pm2N}=A_n,\quad G_{n\pm2N}=G_n.
\label{period}\ea
Onsager \cite{Onsager} derived the following commutation rules:
\be
[A_j,A_k]=4G_{j-k},\quad [G_m,A_l]=2A_{l+m}-2A_{l-m},\quad
[G_j,G_k]=0.
\label{OnsagerAlgebra}
\ee
From these we also have
\be
[A_j,[A_j,[A_j,A_k]]]=16[A_j,A_k],\quad [A_j,[A_j,G_k]]=16G_k,
\ee
compare \cite{DolanGrady}.

We can expand the algebra introducing \cite{JhaValatin}
\be
A^{\alpha\beta}_n=\frac12\sum_{j=1}^N \sigma^\alpha_j
\bigg(\prod_{k=j+1}^{j+n-1}\sigma^z_k\bigg)\sigma^\beta_{j+n},
\quad \alpha,\beta=x,y,
\ee
so that Onsager's $A_n=2A^{xx}_n$, $A_{-n}=2A^{yy}_n$,
$G_n=\mathrm{i}A^{(xy)}_n\equiv\mathrm{i}(A^{xy}_n+A^{yx}_n)$
are recovered, while the commuting \cite{Jha}
$A^{\{xy\}}_n\equiv A^{xy}_n-A^{yx}_n$,
$[A^{\{xy\}}_n,A^{\alpha\beta}_l]=0$, has been added.

If we look at the Jordan--Wigner transform (\ref{Gamma}),
we would expect the $\Gamma_j$ not to be periodic mod $2N$,
but periodic mod $4N$, i.e.,
\be
\Gamma_{j\pm 2N}=P\Gamma_j,\quad \Gamma_{j\pm 4N}=\Gamma_j,\quad P=
\prod_{k=1}^N(-\mathrm{i}\Gamma_{2k-1}\Gamma_{2k}),
\label{Gamma2}\ee
with $P$ given in (\ref{P}).\footnote{This $P$ is called ${}^\dagger\mathbf{U}$ in eq.~(36) of \cite{Kaufman}, with $\mathbf{U}$
from the text above (14) there, not to be confused with $\mathbf{U}$
defined differently in (32) of \cite{Kaufman}.}
To see what this implies, let us evaluate, for
$1\le j,k\le N$,
\ba
\sigma^x_j\bigg(\prod_{k=j+1}^{N+l-1}\sigma^z_k\bigg)
\sigma^x_{N+l}&=&
\sigma^x_j\bigg(\prod_{k=1}^j\sigma^z_k\bigg)
\bigg(\prod_{k=1}^N\sigma^z_k\bigg)
\bigg(\prod_{k=1}^{l-1}\sigma^z_k\bigg)\sigma^x_l\nonumber\\
&=&\bigg(\prod_{k=1}^{j-1}\sigma^z_k\bigg)\sigma^x_j\sigma^z_j
P\bigg(\prod_{k=1}^{l-1}\sigma^z_k\bigg)\sigma^x_l\nonumber\\
&=&\Gamma_{2j-1}(-\mathrm{i}\Gamma_{2j-1}\Gamma_{2j})
P\Gamma_{2l-1}\nonumber\\
&=&-\mathrm{i}\Gamma_{2j}P\Gamma_{2l-1}=
-\mathrm{i}\Gamma_{2j}\Gamma_{2N+2l-1}.
\ea
Hence, with the identification (\ref{Gamma2}), we find
\ba
&&A_n=-\mathrm{i}\sum_{j=1}^N\Gamma_{2j}\Gamma_{2j+2n-1},\\
&&G_n=\mathrm{i}\sum_{j=1}^N(\Gamma_{2j-1}\Gamma_{2j+2n-1}
-\Gamma_{2j}\Gamma_{2j+2n}),\\
&&A^{\{xy\}}_n=-\mathrm{i}\sum_{j=1}^N(\Gamma_{2j-1}\Gamma_{2j+2n-1}
+\Gamma_{2j}\Gamma_{2j+2n}).
\ea
Note that not all the terms are quadratic in fermion operators
due to the identification (\ref{Gamma2}). We will need to follow
Onsager \cite{Onsager} and Kaufman \cite{Kaufman} and split the
state space into a direct sum of even and odd states, corresponding
to eigenvalue of $P$ being $+1$ or $-1$. Each $\Gamma_j$ changes
an even state into an odd state and vice versa. The $A_n$ and $G_n$
act on the odd sector as quadratic in fermions with cyclic boundary
conditions; on the even sector they acquire anticyclic boundary
conditions.\footnote{Sometimes the sector with periodic boundary
conditions is named after Ramond \cite{Ramond} and the one with
antiperiodic boundary conditions after Neveu and Schwarz
\cite{NeveuSchwarz} even though these string theory papers were
written more than two decades later.} In general,
%For example, for $1\le n<N$,
\be
A_n=P_-\,A_n^{(\mathrm{c})}+P_+\,A_n^{(\mathrm{ac})},
%A_n=\frac12(1-P)A_n^{(\mathrm{c})}+\frac12(1+P)A_n^{(\mathrm{ac})},
\ee
with cyclic and anticyclic versions of $A_n$ and projection
operators $P_\pm$,
\be
P_\pm\equiv\frac12(1\pm P),\quad (P_\pm)^2=P_\pm,\quad
P_-+P_+=1,\quad P_-\,P_+=0.
\label{projection}
\ee
Because of (\ref{period}), we only have to consider
the following cases: When $0<n<N$,
\ba
&&A_n^{(\mathrm{c})}=
-\mathrm{i}\sum_{j=1}^{N-n}\Gamma_{2j}\Gamma_{2j+2n-1}
-\mathrm{i}\sum_{j=N-n+1}^N\Gamma_{2j}\Gamma_{2j+2n-2N-1},\\
&&A_n^{(\mathrm{ac})}=
-\mathrm{i}\sum_{j=1}^{N-n}\Gamma_{2j}\Gamma_{2j+2n-1}
+\mathrm{i}\sum_{j=N-n+1}^N\Gamma_{2j}\Gamma_{2j+2n-2N-1}.
\ea
When $-N<n<0$,
\ba
&&A_n^{(\mathrm{c})}=
-\mathrm{i}\sum_{j=1}^{-n}\Gamma_{2j}\Gamma_{2j+2n+2N-1}
-\mathrm{i}\sum_{j=-n+1}^N\Gamma_{2j}\Gamma_{2j+2n-1},\\
&&A_n^{(\mathrm{ac})}=
+\mathrm{i}\sum_{j=1}^{-n}\Gamma_{2j}\Gamma_{2j+2n+2N-1}
-\mathrm{i}\sum_{j=-n+1}^N\Gamma_{2j}\Gamma_{2j+2n-1}.
\ea
Finally,
\ba
A_0^{(\mathrm{c})}=A_0,\quad A_0^{(\mathrm{ac})}=0,\quad
A_N^{(\mathrm{c})}=A_0,\quad A_N^{(\mathrm{ac})}=-A_0.
\ea

If the Hamiltonian $\mch$ is a linear combination of $A_n$'s
with periodic boundary conditions, the partition function
$Z=\mbox{Tr}\,\mathrm{e}^{-\beta\mch}$ can be rewritten as
\be
Z=
\frac12\mbox{Tr}\mathrm{e}^{-\beta\mch^{\mathrm{(c)}}}+
\frac12\mbox{Tr}\mathrm{e}^{-\beta\mch^{\mathrm{(ac)}}}-
\frac12\mbox{Tr}P\mathrm{e}^{-\beta\mch^{\mathrm{(c)}}}+
\frac12\mbox{Tr}P\mathrm{e}^{-\beta\mch^{\mathrm{(ac)}}},
\ee
using $P_-+P_+=1$ (\ref{projection}). In the limit $N\to\infty$ one
can show that the first two terms are asymptotically equal and
infinitely larger than the other two terms
\cite{Katsura,Kaufman,SiskensMazur},
so that we can replace $Z$ by
$\mbox{Tr}\,\mathrm{e}^{-\beta\mch^{\mathrm{(c)}}}$ or by
$\mbox{Tr}\,\mathrm{e}^{-\beta\mch^{\mathrm{(ac)}}}$. Similarly,
in equal-time correlations we can also replace $\mch$
by $\mch^{\mathrm{(c)}}$ or $\mch^{\mathrm{(ac)}}$,
following for example \cite{SiskensMazur}.

However, this method does not work for time-dependent correlations of odd
operators like $\sigma^x_j$. If one wants to keep translation invariance,
one has two options: Following Cheng and Wu \cite{MBA,ChengWu}, one can
asymptotically construct the square of the correlation by
doubling the number of spins in the correlation, so that one only has
even combinations. This leads to infinite (block-Toeplitz) determinants. Otherwise one gets expressions with both $\mch^{\mathrm{(c)}}$ and
$\mch^{\mathrm{(ac)}}$ in it \cite{PC,CvDS,CapelSiskens}.

%%%%%%%%%%%%%%%%%%%%%%%%%%%%%%%%%%%%%%%%%%%%%%%
\section{Conclusion}

We have expressed the pair correlations of the $n$-cluster model,
discussed in \cite{ZG}, explicitly in terms of those of the standard
Ising chain in transverse field. These latter correlations are known
in great detail, especially in the ground state \cite{Pfeuty,AP}.
In particular, one can use the 2-dimensional Ising recurrence
relations and asymptotic results of section 2 in \cite{AP}, together
with (60)--(63) there to obtain very accurate results for $X(n)$ and
$X^{\ast}(n)$ in the ground state. Then the $Y(n)$ and $Y^{\ast}(n)$
can be obtained from (37), (38), (71) and (72) for $t=0$ in \cite{AP}.

We also factorized the pair correlation of two cluster operators
of the form (24) in \cite{ZG} as a product of $n+1$ factors of the
form $X(n)$, which can also be given to great accuracy using \cite{AP}.

It should be noted that the time-dependent $xx$, $xy$, $yx$ and $yy$ correlations have factorizations similar to those in (\ref{xy0}),
(\ref{xy}) and (\ref{zz0}), so that results from \cite{AP} and references
cited there can be used.

Next, we note that we can derive similar factorizations for
cluster models with Hamiltonians of the form
$\mch=\lambda A_n+\mu A_m$, with $n$ and $m$ arbitrary. Then
there is a factorization into $|m-n|$ Ising chains in transverse
field similar to (\ref{factored}). As an
example one may consider the zero-field cluster model with
Hamiltonian of the form $\mch=-J_x A_{n+1}-J_y A_{-n-1}$, or
\be
\mch=-J_x\sum_{j=1}^N\sigma^x_j
\bigg(\prod_{k=j+1}^{j+n}\sigma^z_k\bigg)\sigma^x_{j+n+1}-
J_y\sum_{j=1}^N\sigma^y_j\bigg(\prod_{k=j+1}^{j+n}\sigma^z_k\bigg)
\sigma^y_{j+n+1},
\ee
with $N\equiv(n+1)N_1$. In this case there are $2(n+1)$ factors.
We leave that as an exercise for the interested reader.

Factorizations can also occur if there are more terms in the
Hamiltonian, which is assumed to be of the form
$\mch=\sum_{k=0}^r\lambda_kA_{n+km}$. For example,
\be
\mch=-\sum_{j=1}^N \bigg[J_x\sigma^x_j
\bigg(\prod_{k=j+1}^{j+n}\sigma^z_k\bigg)\sigma^x_{j+n+1}+
J_y\sigma^y_j\bigg(\prod_{k=j+1}^{j+n}\sigma^z_k\bigg)
\sigma^y_{j+n+1}+B\sigma^z_j\bigg],
\label{XYs}
\ee
or $\mch=-J_x A_{n+1}-J_y A_{-n-1}+BA_0$. Now there are $n+1$
factors corresponding to different anisotropic XY-chains in
transverse field. The correlation functions can all be factored
in full detail like is done in section 2. It should be remarked
that Minami \cite{Minami2} very recently briefly noted this
decoupling of Hamiltonian (\ref{XYs}), without discussing the
factorization of the correlation functions. In two papers
\cite{Minami2,Minami1} he also presented a general construction of
models equivalent to the Ising chain in transverse field, in
essence giving representations of the Temperley--Lieb algebra \cite{TL},
\be
e_i^{\,2}=\sqrt{2}\,e_i^{\vphantom{2}},\quad
e_i^{\vphantom{2}}e_{i\pm1}^{\vphantom{2}}e_i^{\vphantom{2}}=e_i^{\vphantom{2}},\quad
[e_i^{\vphantom{2}},e_j^{\vphantom{2}}]=0\mbox{ if }|i-j|>2,
\ee
identifying $\eta_i=1-\sqrt{2}\,e_i$ in \cite{Minami2,Minami1}.

Finally, so far we have taken all interaction constants to be
uniform. If we relax that, we no longer have the Onsager algebra
(\ref{OnsagerAlgebra}), but a larger algebra instead. However all
factorizations still work the same way.

%%%%%%%%%%%%%%%%%%%%%%%%%%%%%%%%%%%%%%%%%%%%%%%

\end{document}